\documentstyle[amssymb,aps]{revtex}
\begin{document}
\draft
\title{Broadband Dielectric Spectroscopy on Glass-Forming Propylene Carbonate}
\author{U. Schneider, P. Lunkenheimer, R. Brand, and A. Loidl}
\address{Experimentalphysik V, Universit\"{a}t Augsburg, D-86135 Augsburg, Germany}
\date{}
\maketitle

\begin{abstract}
Dielectric spectroscopy covering more than 18 decades of frequency has been
performed on propylene carbonate in its liquid and supercooled-liquid state.
Using quasi-optic submillimeter and far-infrared spectroscopy the dielectric
response was investigated up to frequencies well into the microscopic
regime. We discuss the $\alpha $-process whose characteristic timescale is
observed over 14 decades of frequency and the excess wing showing up at
frequencies some three decades above the peak frequency. Special attention
is given to the high-frequency response of the dielectric loss in the
crossover regime between $\alpha $-peak and boson-peak. Similar to our
previous results in other glass forming materials we find evidence for
additional processes in the crossover regime. However, significant
differences concerning the spectral form at high frequencies are found. We
compare our results to the susceptibilities obtained from light scattering
and to the predictions of various models of the glass transition.
\end{abstract}


%\begin{multicols}{2}
%\columnseprule 0pt
%\narrowtext

\section{INTRODUCTION}

Despite a long history of research on the glass transition \cite{hist}, this
phenomenon is still commonly regarded as an unresolved problem. In recent
years a variety of new theoretical and phenomenological approaches of the
glass transition (e.g., \cite
{mct,mctrev,mctext,Ngai,Chamb,Kiv,Nagscal,Nagsus}) stimulated new
experimental investigations of the dynamic response of glass-forming liquids
(see, e.g. \cite
{Rich,nsrev,Cum,procs,Wutt,Du,PimPRE,Funke,Lunkigly,Lunkiorl,Lunkickn}).
Here dielectric spectroscopy has played an important role, mainly due to the
broad dynamic range accessible with this method (e.g., \cite
{Nagscal,Nagsus,PimPRE,Funke,Lunkigly,Lunkiorl,Lunkickn,diel,wing}). Spectra
of the dielectric loss $\varepsilon ^{\prime \prime }$ show a variety of
features, the microscopic origin of most of them being still controversially
discussed. Most prominent is the $\alpha $-peak associated with the well
known $\alpha $- or structural-relaxation process. The comparison of the
non-trivial temperature dependence of the $\alpha $-relaxation time with
theoretical predictions states an important test for any model of the
glass-transition. Some decades above the $\alpha $-peak frequency $\nu _{p}$%
, an excess wing (also called ''high-frequency wing'' or ''tail'') shows up
as high-frequency excess contribution to the power law $\nu ^{-\beta }$,
commonly found at $\nu >\nu _{p}$. This excess wing seems to be a universal
feature of glass-forming liquids as can be deduced from the scaling behavior
found by Nagel and coworkers \cite{Nagscal}. Based on this universal scaling
of $\alpha $-relaxation and excess wing, Nagel and coworkers \cite{Nagsus}
proposed a divergence of the static susceptibility which essentially implies
a constant loss behavior at high frequencies and low temperatures as has
early been predicted by Wong and Angell \cite{Wong}. The interest in even
higher frequencies, in the GHz-THz region, was mainly stimulated by the mode
coupling theory (MCT) \cite{mct,mctrev,mctext}, which explains the glass
transition in terms of a dynamic phase transition at a critical temperature $%
T_{c\text{ }}$significantly above the glass temperature $T_{g}$. From
neutron- and light-scattering experiments it is well known that in the THz
region the so-called boson peak shows up in the imaginary part of the
susceptibility of glass-forming liquids. For the transition region between $%
\alpha $-process and boson peak, MCT predicts an additional contribution,
now commonly termed fast $\beta $-relaxation. But also other models predict
fast processes, the most prominent being the coupling model by Ngai and
coworkers \cite{Ngai}. This transition region was mainly investigated by
neutron and light scattering and indeed indications for additional fast
processes were found \cite{nsrev,Cum,Wutt,Du}. The scattering results were
described within the framework of MCT and partly good agreement with the
theoretical results was obtained. However, the applicability of the MCT for
glass-forming liquids, especially at low temperatures near $T_{g}$, is still
a matter of controversy. Dielectric data were scarce in the relevant
high-frequency region as for classical dielectric methods the region above
GHz is not accessible. Only recently, our group was able to obtain
continuous dielectric spectra on glass-forming liquids extending well into
the submillimeter wavelength region \cite
{Lunkigly,Lunkiorl,Lunkickn,Lunkikyo,Lunkibos,Schn}. The existence of a
broad minimum in $\varepsilon ^{\prime \prime }(\nu )$ was found which
cannot be explained by a simple crossover from the structural ($\alpha $-)
relaxation to the far-infrared (FIR) vibrational response. Similar to the
scattering results, clear indications for additional fast processes
prevailing in this transition region were found. For glass-forming glycerol,
by combining classical dielectric spectroscopy, coaxial transmission,
quasi-optic submillimeter and far-infrared techniques, spectra covering 18
decades of frequency and extending well into the THz range were obtained 
\cite{Schn}. This allowed for the observation of the complete dynamic
response including the boson peak. In addition, dielectric spectra up to 380
GHz were obtained in [Ca(NO$_{3}$)$_{2}$]$_{0.4}$[KNO$_{3}$]$_{0.6}$ (CKN)
and [Ca(NO$_{3}$)$_{2}$]$_{0.4}$[RbNO$_{3}$]$_{0.6}$ (CRN) \cite
{Lunkiorl,Lunkickn,Lunkikyo}. Both are ionically conducting, molten-salt
glass formers with relatively high fragility \cite{strong,m}, $m\approx 90$ 
\cite{4glass}. In contrast, glycerol is a rather strong hydrogen-bonded
glass-former with a fragility parameter of $m\approx 53$ \cite{rol}. In many
respects, especially concerning the high-frequency response near the boson
peak and the agreement of the dielectric data with MCT predictions and with
the scattering results, glycerol and the molten salts behave qualitatively
different. In order to clarify the origin of these differences and in the
light of the dependence of the boson peak contribution on fragility \cite
{Sok}, it seemed of interest to investigate a fragile, but molecular
glass-former. For this purpose we have chosen propylene carbonate (PC) which
like glycerol is a molecular glass former, but can be characterized as a
fragile ($m\approx 104$,\cite{rol}) van der Waals liquid. Preliminary
results on PC in a somewhat restricted frequency range have been published
earlier \cite{Lunkiorl,Lunkikyo,Lunkibos}. In the present paper we present
extended results covering the complete dynamic range including the boson
peak. We address all features seen in the dielectric spectra, i.e. the $%
\alpha $-response, the excess wing, the fast $\beta $-relaxation region and
the boson peak. The results are compared with a variety of theoretical
predictions and with the findings from other experimental methods.

\section{EXPERIMENTAL DETAILS}

The measurements presented in this article have been performed using a
variety of different techniques to cover nearly 20 decades in frequency. The
lowest frequencies ($10^{-6}\lesssim \nu \lesssim $ $10^{3}$\thinspace Hz)
were investigated in the time-domain using a spectrometer that is based on a
design described by Mopsik \cite{Mops}. The complex dielectric constant was
calculated from the measured response function via a Fourier transformation.
The autobalance bridges HP4284 and HP4285 were used in the range $20$ Hz $%
\leq \nu \leq 20$ MHz. For the radio-frequency and microwave range ($1$ MHz $%
\leq \nu \leq 10$ GHz) a reflectometric technique was employed \cite{jappl}
using the HP4191 and HP4291 impedance analyzers and the HP8510 network
analyzer. In addition, at frequencies $100$ MHz $\leq \nu \leq 30$ GHz data
were taken in transmission with the HP8510 network analyzer using 7mm
coaxial lines of various lengths (between 10 and 300 mm) filled with the
sample material. Closed-cycle refrigerators, N$_{2}$-, and He-cryostats have
been used to cover the relevant temperature ranges.

At frequencies $40$ GHz $\leq \nu \leq 1.2$ THz a quasi-optical
submillimeter spectrometer was used with an experimental arrangement similar
to a Mach-Zehnder interferometer \cite{Volkov}. This setup allows for
measuring the frequency dependence of both the transmission and the phase
shift of a monochromatic electromagnetic beam through the sample. The
frequency range up to 1.2\thinspace THz is covered continuously by 10
tunable narrow-band backward-wave oscillators (BWOs). The signal was
detected by a Golay cell or a pumped He bolometer and amplified using
lock-in techniques. The liquid was put in specially designed cells made of
polished stainless steel with thin plane-parallel quartz windows; depending
on the range of frequency and temperature the thickness of the sample cell
was between 1 mm and 30 mm. The sample cell was placed in a home-made
cryostat and cooled by a continuous flow of nitrogen gas. The data were
analyzed using optical formulae for multilayer interference \cite{Born} with
the known thickness and optical parameters of the windows in order to get
the real and imaginary part of the dielectric constant of the sample as a
function of frequency at various temperatures. At higher frequencies
(600\thinspace GHz $-$ 3\thinspace THz) the samples were investigated with a
Fourier-transform infrared spec\-tro\-meter covering a regime from
450\thinspace GHz to 10\thinspace THz. This set-up allows the measurement of
the transmission or reflection only, whereas the phase shift caused by the
sample cannot be determined.

To cover the complete frequency range, a single $\varepsilon ^{\prime \prime
}(\nu )$ curve at a given temperature is superimposed using results from the
various setups. The time-domain results are obtained in arbitrary units and
scaled by one factor to match the autobalance bridge results. For the
measurements at $1$ GHz $\leq \nu \leq 40$ \thinspace GHz there are some
uncertainties of the absolute values originating from an ill-defined
geometry of the samples or parasitic elements. These results have been
partly shifted by one scaling factor per measurement series which led to a
good match at both the low and the high frequency side. Typically, shifting
factors of less than $20\%$ had to be applied to obtain smooth curves. The
submillimeter-frequency results and, at losses $\varepsilon ^{\prime \prime
}\gtrsim 1$, also the coaxial transmission measurements provide excellent
absolute values of $\varepsilon ^{\prime }$ and $\varepsilon ^{\prime \prime
}$. The values for $\varepsilon ^{\prime \prime }$ from the infrared
measurements were calculated consistently with the Kramers-Kronig relation
assuming a reasonable behavior of the dielectric constant $\varepsilon
^{\prime }$ which is almost constant in this frequency regime. In this
latter case, the error bars for $\varepsilon ^{\prime \prime }$ were
estimated by variation of the assumed $\varepsilon ^{\prime }$ behavior.

As sample material, propylene carbonate ($T_{g}=159$ K \cite{AngPC}, $%
T_{m}=218$ K) with a purity of $99.7\%$, was used. To avoid crystallization
polished sample holders had to be used in the various experiments. However,
occasional crystallization, mainly at temperatures around $180$ K could not
be avoided completely. Therefore some measurements were performed after
heating up to $T_{m}$ and direct cooling to each measurement temperature
with relatively fast cooling rates ($1$ K/min).

\section{RESULTS}

\subsection{$\protect\alpha $-relaxation and excess wing}

Figures \ref{eps1} and \ref{eps2} show $\varepsilon ^{\prime }(\nu )$ and $%
\varepsilon ^{\prime \prime }(\nu )$ for various temperatures in the whole
accessible frequency range. $\varepsilon ^{\prime \prime }(\nu )$ (Fig. \ref
{eps2}) exhibits the typical asymmetrically shaped $\alpha $-relaxation
peaks shifting through the frequency window with temperature. They are
accompanied by relaxation steps in $\varepsilon ^{\prime }(\nu )$ as seen in
Fig. \ref{eps1}. In most respects the data agree well with the results of
earlier dielectric investigations of PC \cite
{Johari,Payne,Masood,Huck,Barthel,Schoen,AngPC,BohmPC,Le} which were
restricted to smaller frequency and temperature ranges. However, some
differences show up in the absolute values (see below). The solid lines in
Figs. \ref{eps1} and \ref{eps2} are fits of the $\alpha $-relaxation region
with the empirical Cole-Davidson function \cite{CD}, $\varepsilon ^{\ast
}=\varepsilon _{\infty }+(\varepsilon _{s}-\varepsilon _{\infty })/(1+i2\pi
\nu \tau _{CD})^{\beta _{CD}}$, performed simultaneously for real and
imaginary part. $\varepsilon _{s}$ and $\varepsilon _{\infty }$ denote the
low- and high-frequency limit of the dielectric constant, respectively. A
good fit of the peak region was achieved. Very close to the maximum the loss
curves can also be described by the Fourier transform of the
Kohlrausch-Williams-Watts (KWW) function \cite{KWW}, $\Phi =\Phi _{0}\exp
[-(t/\tau _{KWW})^{\beta _{KWW}}]$, (dotted lines, shown for 173 K, only).
For all temperatures investigated the KWW fit is of lower quality than the
CD fit. This can be ascribed to the fact that there is a significant
difference of CD and KWW response concerning the loss-peak region. For the
KWW response the curvature near the peak is retained somewhat further above
the peak frequency and a power law $\varepsilon ^{\prime \prime }\thicksim
\nu ^{-\beta _{KWW}}$ is approached at significantly higher frequencies
only. Therefore, in most cases the exponent $\beta _{KWW}$ obtained from the
KWW fits differs significantly from the power law exponent $\beta $ actually
observed for $\nu >\nu _{p}$. In contrast, for the CD fits $\beta
_{CD}=\beta $ is found. Despite the KWW function is more widely used
nowadays, in the authors' experience dielectric loss data in glass-forming
materials are often described much better by the CD function \cite
{wing,Lunkibos,Schn,Brand}. The inset of Fig. \ref{eps2} shows the
temperature dependence of the frequency $\nu _{\tau }=1/(2\pi <\tau >)$
which is virtually identical to the peak frequency $\nu _{p}$. Here $<\tau >$
denotes the mean relaxation time \cite{Boett} calculated from $<\tau
>_{CD}=\beta _{CD}\tau _{CD}$ for the CD function and $<\tau >_{KWW}=\tau
_{KWW}/\beta _{KWW}\times \Gamma (1/\beta _{KWW})$ ($\Gamma $ denoting the
Gamma function) for the KWW function. The results from the CD (circles) and
the KWW fits (pluses) agree perfectly well and are in accord with previously
published data \cite{Johari,Huck,Barthel,Schoen,AngPC,Stick2}. $\nu _{\tau
}(T)$ shows the well known deviations from thermally activated behavior,
typical for fragile glass-formers. It can be parameterized using the
Vogel-Fulcher-Tammann (VFT) equation \cite{VFT} (line in the inset of Fig. 
\ref{eps2}; see section \ref{discalp}).

In Figure \ref{betchi}(a) the relaxation strengths $\Delta \varepsilon =$ $%
\varepsilon _{s}-\varepsilon _{\infty }$ obtained from the CD and KWW fits
are shown. One has to be aware that $\Delta \varepsilon $ resulting from the
fits is somewhat smaller that $\Delta \varepsilon $ read off from the $%
\varepsilon ^{\prime }(\nu )$ data of Fig. \ref{eps1}. As seen in Fig. \ref
{eps1}, the fits overestimate $\varepsilon _{\infty }$ because there is an
additional decrease of $\varepsilon ^{\prime }(\nu )$ corresponding to the
excess wing contribution (see below) which is not taken into account by the
CD fits. Consequently $\Delta \varepsilon $ shown in Fig. \ref{betchi}(a)
presents the relaxation strength of the $\alpha $-process alone, without the
excess wing contribution. $\Delta \varepsilon $ decreases monotonically with
temperature which is in accord with previous reports \cite
{Payne,Schoen,AngPC}. However, the decrease found in the present work is
somewhat less steep than reported earlier. Unfortunately, in \cite{Schoen}
where a rather broad temperature and frequency range was investigated no
absolute values of $\Delta \varepsilon $ were reported. The results from the
high frequency measurements of Payne and Theodorou \cite{Payne} are
compatible with our data at $T>273$ K except for their lowest temperature ($%
195$ K) where a much higher value ($\Delta \varepsilon \approx 86$) was
found. However, the relaxation time at $195$ K reported by these authors
deviates by more than one decade from that found in the present and other
works \cite{Schoen,Stick2} which sheds some doubt upon the significance of
the results at this temperature. Also from the results of Huck {\it et al.} 
\cite{Huck} and Angell {\it et al. }\cite{AngPC}{\it \ }near $170$ K a much
higher magnitude of $\Delta \varepsilon $ ($>100$) than determined here can
be read off. The determination of the absolute values of $\Delta \varepsilon 
$ is a difficult task as it depends on the shifting factors necessary to
match results from different experimental set-ups and on the subtraction of
possible stray capacitances inherent to some methods. We believe, that due
to the broad overlap of the frequency ranges of the various devices used in
the present work, the reproducibility of results with different sample
holders, and the simultaneous fitting of $\varepsilon ^{\prime }(\nu )$ and $%
\varepsilon ^{\prime \prime }(\nu )$, Fig. \ref{betchi}(a) gives a good
estimate of the absolute value of $\Delta \varepsilon $.

Figure \ref{betchi}(b) shows the width parameter obtained from both fitting
procedures. Both, $\beta _{CD}$ and $\beta _{KWW}$ increase nearly linearly
with temperature up to about $200$ K. Above this temperature a tendency to
saturate at a value below unity is observed. This is in contrast to \cite
{Schoen} where $\beta _{CD}(T)$ was reported to saturate at unity for high
temperatures. This discrepancy may be due to the smaller frequency range
available in \cite{Schoen} which for high temperatures leads to a
restriction of the data at the high-frequency side of the loss peaks, that
is essential for the determination of $\beta _{CD}$. Unfortunately, no $%
\varepsilon ^{\prime \prime }(\nu )$ data of PC are shown in \cite{Schoen}.

At frequencies about 2-3 decades above $\nu _{p}$ deviations of $\varepsilon
^{\prime \prime }(\nu )$ from the CD-fits show up. For low temperatures
these deviations can be described as second power law, $\varepsilon ^{\prime
\prime }\thicksim \nu ^{-b}$ with $b<\beta $, in addition to the power law $%
\varepsilon ^{\prime \prime }\thicksim \nu ^{-\beta }$ constituting the high
frequency flank of the $\alpha $-peak. The exponent $b$ decreases with
decreasing temperature as found previously for PC and other glass-forming
materials \cite{Le}. The excess wing is accompanied by a decrease in $%
\varepsilon ^{\prime }(\nu )$ as mentioned above. It shows up as a somewhat
smoother rounding of the $\varepsilon ^{\prime }(\nu )$ curves (compared,
e.g., to the CD behavior) when approaching $\varepsilon _{\infty }$ (Fig. 
\ref{eps1}). For $T=193$ K the exponent $b$ has reached a value almost
identical with $\beta _{CD}$ (Fig. \ref{eps2}) and for $203$ K the excess
wing seems to have merged with the $\alpha $-peak. For $T\geq $ $203$ K the
deviations of the experimental data from the CD-fits are due to the fast
dynamic processes described in the following section.

\subsection{Fast dynamics}

\label{resfast}At high frequencies, succeeding the excess wing for low and
the $\alpha $-peak for high temperatures, $\varepsilon ^{\prime \prime }(\nu
)$ exhibits a smooth transition into a shallow minimum. Such a minimum was
earlier observed in glycerol \cite{Lunkigly,Lunkiorl,Lunkibos,Schn}, Salol 
\cite{Lunkiorl,Lunkibos}, the molten salts CKN \cite
{PimPRE,Funke,Lunkiorl,Lunkickn,Lunkikyo} and CRN \cite{Lunkickn}, and the
plastic crystals cyclo-octanol \cite{Brand,Robdip} and ortho-carborane \cite
{Robdip}. With increasing temperature, the amplitude $\varepsilon _{\min }$
and frequency position $\nu _{\min }$ of the minimum increases. For room
temperature the measurements have been extended into the FIR region. Here no
minimum is observed, but a shoulder shows up near 1 THz. This shoulder is
indicative of the so-called boson peak known mainly from neutron- and
light-scattering measurements \cite{nsrev,Cum,Wutt,Du}. Indeed, for PC the
boson peak, determined from light-scattering measurements \cite{Du}, is
located just above 1 THz. As the peak frequency and amplitude is only weakly
temperature dependent, it becomes obvious that the high-frequency flank of
the $\varepsilon ^{\prime \prime }(\nu )$-minimum is identical to the
low-frequency flank of the boson peak. The frequency dependence in this
region can be approximately described by a power law $\varepsilon ^{\prime
\prime }\thicksim \nu ^{a}$ . The exponent $a$ increases with decreasing
temperature and seems to approach a linear behavior as indicated by the
dash-dotted line in Fig. \ref{eps2}. In $\varepsilon ^{\prime }(\nu )$ for
these frequencies the onset of a relaxation step corresponding to the boson
peak can be seen (Fig. \ref{eps1}).

In Fig. \ref{vgl} we compare $\varepsilon ^{\prime \prime }(\nu )$ with $%
\chi ^{\prime \prime }(\nu )$ calculated from the light scattering results
on PC \cite{Du}. As the light scattering results give no information on the
absolute values of $\varepsilon ^{\prime \prime }$, the datasets have been
scaled to yield a comparable height of the boson peak. While the dielectric
results in PC are qualitatively similar to the light scattering results
(succession of $\alpha $-peak, minimum, and boson peak), quantitative
differences show up: The $\alpha $-peak is located at a significantly lower
frequency and the ratio of the amplitudes of $\alpha $- and boson peak is
higher for the dielectric measurements. The latter behavior was also found
in our experiments on glycerol \cite{Lunkibos,Schn} and Salol \cite{Lunkibos}%
, and in molecular dynamics simulations of ortho-terphenyl \cite{Wahn} and
of a system of rigid diatomic molecules \cite{Schsim}. The results from a
very recent neutron scattering study of PC \cite{Ohl} indicate an even
smaller ratio of $\alpha $- and boson peak amplitude than for light
scattering. Clearly the position of the minimum differs between the
different methods, also in accordance with results on other systems \cite
{Lunkibos,Schn,Schsim}. In contrast, the increase towards the boson peak
exhibits almost identical power laws for the dielectric and light scattering
measurements.

\section{DISCUSSION}

\subsection{$\protect\alpha $-relaxation}

\label{discalp}The most significant result of an analysis of the $\alpha $%
-relaxation process is the temperature dependence of the $\alpha $%
-relaxation time which is compared to various predictions in Fig. \ref{tau}.
Figure \ref{tau}(a) shows the same fit with the VFT equation \cite{VFT},

\begin{equation}
\nu _{\tau }=\nu _{0}\exp \left( \frac{-DT_{VF}}{T-T_{V\!F}}\right)
\label{eqVFT}
\end{equation}
as in the inset of Fig. \ref{eps2}. A VFT temperature $T_{V\!F}=132$ K and a
strength parameter $D=6.6\;$were determined, the latter characterizing PC as
fragile glass former \cite{strong}. While the VFT equation is primarily an
empirical description, a theoretical foundation was given in various models,
e.g. the Adam-Gibbs theory \cite{AG} or the free volume theory \cite{FV}.

In Fig. \ref{tau}(a) it becomes obvious that, above a crossover temperature $%
T_{A}\approx 220$ K, deviations from VFT behavior show up, similar to those
seen in earlier work \cite{Schoen,AngPC,Stick2}. In \cite{Schoen} and \cite
{AngPC} a transition to thermally activated behavior was suggested as
demonstrated by the dashed line in Fig. \ref{tau}(a). $T_{A}$ was
interpreted \cite{AngPC} as temperature below which the potential energy
landscape becomes important. However, one has to state that also alternative
descriptions are possible, e.g. using VFT behavior at high and a thermally
activated behavior at low temperatures or a combination of VFT functions 
\cite{cumcom}. In addition, Stickel et al. \cite{Stick2} performed an
analysis of relaxation-time data on PC extending to temperatures above room
temperature using a temperature-derivative method \cite{Stick1}. They
reported a transition between two VFT laws at about $200$ K and a second
transition to thermally activated behavior above about $300$ K. Overall, as
demonstrated in Fig. \ref{tau}(a), the three-parameter VFT function is not
able to describe the data in the whole temperature range and a crossover to
a different behavior has to be assumed to overcome this difficulty.

A four-parameter function for the description of $\nu _{\tau }(T)$ is the
outcome of the extended free volume (EFV) theory \cite{EFV}. In the EFV
theory the supercooled liquid is assumed to be divided into liquid-like and
solid-like regions. At a temperature $T_{0}$ a transition from percolating
(fluid) to isolated liquid-like regions (solid glass state) occurs. By
addressing the entropy of the glass-forming system and making use of
percolation theory the following result for $\nu _{\tau }(T)$ was obtained:

\begin{equation}
\log _{10}(\nu _{\tau })=-A-\frac{B}{T-T_{0}+[(T-T_{0})^{2}+CT]^{\frac{1}{2}}%
}  \label{eqEFV}
\end{equation}
Good fits with Eq. (\ref{eqEFV}) for a variety of glass-formers were
reported \cite{Cum,EFV}. Figure \ref{tau}(b) shows a fit of $\nu _{\tau }(T)$
of PC with Eq. (\ref{eqEFV}) and $T_{0}=162$ K. Indeed, the EFV theory gives
a very good description of the present data and $T_{0}$ is almost identical
to the calorimetric $T_{g}\approx 159$ K \cite{AngPC}.

While the VFT and EFV predictions lead to critical temperatures near or
below $T_{g}$, the $T_{c}$ of MCT is predicted to be located well above $%
T_{g}$. The simplest version of MCT, the idealized MCT \cite{mct,mctrev},
predicts a critical behavior of the $\alpha $-relaxation timescale, 
\begin{equation}
\nu _{\tau }\thicksim (T-T_{c})^{\gamma }  \label{eqMCTfp}
\end{equation}
with a critical exponent $\gamma $ that is determined by $\gamma
=1/(2a)+1/(2b)$. Here $a$ and $b$ are the low- and high-frequency power law
exponents of the $\varepsilon ^{\prime \prime }(\nu )$-minimum as will be
explained in detail in section \ref{secdiscfastmin}. The solid line in Fig. 
\ref{tau}(c) is a fit with the three-parameter function, Eq. \ref{eqMCTfp}
with $\gamma =2.72$ and $T_{c}=187$ K fixed to the values obtained from the
analysis of the $\varepsilon ^{\prime \prime }(\nu )$-minimum (sect. \ref
{secdiscfastmin}). Even with all parameters free, very similar values
result, namely $T_{c}=185$ K and $\gamma =2.78$. For the fits, only data
above 200 K were used where a good agreement of data and fit was achieved.
In order to describe the data at lower temperatures a VFT fit can be
employed (dashed line). The observed absence of critical behavior at $T_{c}$
in the experimental data is expected within the extended MCT \cite
{mctrev,mctext}. Here the structural arrest found in idealized MCT for $%
T<T_{c\text{ }}$is avoided by the assumption of thermally activated hopping
processes. MCT also predicts that, at least for $T>T_{c}$, the relaxation
times determined with different experimental methods should follow a common
temperature behavior. Indeed, while there is a difference in the absolute
values of $\tau $ from light scattering \cite{Du} and dielectric
measurements (leading to the different $\alpha $-peak positions in Fig. \ref
{vgl}), which can be understood considering the different correlation
functions of both methods \cite{Schpriv}, they can be transferred into each
other by a temperature independent factor of about 3 (not shown).

Finally, we compare the results to the predictions of the
frustration-limited domain (FLD) model by Kivelson, Tarjus, and coworkers 
\cite{Kiv}. This theory postulates a ''narrowly avoided critical point'' at
a temperature $T^{\ast }$ above the melting point. The theory is based on
the assumption that there is a locally preferred structure (LPS) which,
however, is not able to tile space periodically. Without this geometrical
constraint the system would condense into the LPS at $T^{\ast }$. In real
systems somewhat below $T^{\ast }$, frustration-limited domains with the LPS
are formed. Such a scenario is intuitive for the simple system of spherical
molecules. Here the LPS is an icosahedral short-range order but one cannot
tile space with this structure. By using a phenomenological scaling
approach, a prediction for the temperature dependent relaxation rate was
obtained: 
\begin{eqnarray}
\nu _{p} &=&\nu _{\infty }\exp \left( -\frac{E_{\infty }}{T}\right) \text{
for }T>T^{\ast }  \nonumber \\
\nu _{p} &=&\nu _{\infty }\exp \left[ -\frac{E_{\infty }}{T}-\frac{FT^{\ast }%
}{T}\left( \frac{T^{\ast }-T}{T^{\ast }}\right) ^{8/3}\right] \text{ for }%
T<T^{\ast }  \label{Kiveq}
\end{eqnarray}

Good agreement of this four-parameter function with the results in a variety
of glass-formers was found \cite{Kiv96}. This also is valid for the present
results on PC as demonstrated in Fig. \ref{tau}(d).

Clearly, among the analyses presented in Fig. \ref{tau}, the one employing
the combination of VFT and thermally activated behavior [Fig. \ref{tau}(a)]
seems least reasonable as there is no theoretical base for such a behavior.
As could be expected, the two four-parameter functions (EFV and FLD model)
provide the best descriptions of the data as was also found for a similar
comparison performed for dielectric results on Salol \cite{Cum}. As
mentioned above, the huge deviations seen for the (idealized) MCT approach
could be expected. This setback may be overcome when employing the extended
MCT, which is out of the scope of the present work. An argument in favor of
the MCT-fit is the fact that only one parameter was varied, $T_{c}$ and $%
\gamma $ being fixed as mentioned above.

The MCT also makes distinct predictions for the temperature dependence of
the relaxation strength $\Delta \varepsilon $ and the spectral form of the $%
\alpha $-process \cite{mct,mctrev,mctext}. According to idealized MCT, for $%
T>T_{c}$, the relaxation strength and the spectral form of the $\alpha $%
-process should be temperature independent (time-temperature superposition
principle). In addition, for $T<T_{c}$, $\Delta \varepsilon
=c_{1}+c_{2}[(T_{c}-T)/T_{c}]^{%
%TCIMACRO{\UNICODE[m]{0xbd}}%
%BeginExpansion
{\frac12}%
%EndExpansion
}$ follows from extended MCT. In Fig. \ref{betchi}(a) the solid line was
calculated using the MCT prediction with $T_{c}=187$ K as deduced in section 
\ref{secdiscfastmin}. Due to the scattering of the data no definite
conclusion can be drawn, but at least the data do not contradict MCT. In
contrast Schoenhals {\it et al.} \cite{Schoen} reported clear deviations of
their $\Delta \varepsilon (T)$ (given in arbitrary units) from MCT
predictions \cite{remschoen}. Concerning the predicted temperature
independent spectral form of the $\alpha $-peak for $T>T_{c}$, a tendency to
saturate is indeed seen for $\beta _{CD}(T)$ and $\beta _{KWW}(T)$ in Fig. 
\ref{betchi}, however at temperatures clearly above $T_{c}$, only. But a
reasonable description of the $\alpha $-peaks for $T>T_{c}$ is also possible
with a constant $\beta _{CD}\approx 0.8$. Again this is in contrast to the
statements made in \cite{Schoen}. Overall it is difficult to make a decisive
statement about the validity of MCT in PC from the analysis of the $\alpha $%
-process alone.

\subsection{Excess wing}

The excess wing (also called ''high-frequency wing'' or ''tail'') seems to
be a universal feature of glass-forming liquids \cite{Nagscal,remwing}, but
up to now its microscopic origin remains unclear. A phenomenological
function taking account of the $\alpha $-peak and the wing has been proposed
recently \cite{Kudphen}. In many cases it is possible to describe the $%
\alpha $-peak including the wing using a model of dynamically correlated
domains \cite{Chamb} but for low temperatures and extremely broadband data
deviations show up \cite{Chambrem}.

Nagel and coworkers \cite{Nagscal} found that the $\varepsilon ^{\prime
\prime }(\nu )$-curves for different temperatures and even for different
materials, including the $\alpha $-peak and the wing, can be scaled onto one
master curve by an appropriate choice of the x- and y-axis. This intriguing
scaling behavior strongly suggests a correlation between $\alpha $-process
and the high-frequency wing. During the last years some criticism of the
Nagel-scaling arose concerning its universality \cite{Schoen2,Dendz} and
accuracy \cite{Dendz,Kud} and minor modifications of the original scaling
procedure have been proposed \cite{Dendz}. However, it is still commonly
believed that the Nagel-scaling is of significance for our understanding of
the glass-forming liquids and many efforts have been made to check its
validity in a variety of materials \cite{Brand,Schoen2,Kud,Men,LP}. In Fig. 
\ref{scal} we have applied the modified scaling approach proposed by Dendzik 
{\it et al.} \cite{Dendz}. This procedure was shown to lead to somewhat
better scaling in the $\alpha $-peak region and allows for an unequivocal
determination of the scaling parameters. In addition, the modified procedure
is able to collapse different CD-curves onto one master curve \cite{Dendz}
which is not the case for the original Nagel-scaling \cite{Men}. As
demonstrated in Fig. \ref{eps2}, the CD function gives an excellent
description of the $\alpha $-peak region in $\varepsilon ^{\prime \prime
}(\nu )$. The application of the modified scaling requires the determination
of parameters $\nu _{s}$, $\varepsilon _{s}^{\prime \prime }$, and $\beta $.
Here $\nu _{s}$ and $\varepsilon _{s}^{\prime \prime }$ are defined as
frequency and amplitude of the intersection point of the two power laws $%
\varepsilon ^{\prime \prime }\thicksim \nu $ for $\nu <\nu _{p}$ and $%
\varepsilon ^{\prime \prime }\thicksim \nu ^{-\beta }$ for $\nu >\nu _{p}$.
In Fig. \ref{scal} we show the PC results in addition to glycerol data taken
from \cite{Schn}, scaled according to \cite{Dendz}, i.e. with $x=\left(
1+\beta \right) \log _{10}(\nu /\nu _{s})$ and $y=\log _{10}[\varepsilon
^{\prime \prime }\nu _{s}/(\varepsilon ^{\prime \prime }\nu )]$. Indeed, the 
$\alpha $-peak and the excess wing region can nicely be scaled onto one
master curve for each material. The deviations from this master curve seen
in Fig. \ref{scal} occur in the wing and boson peak region which cannot be
scaled in this way. The scaling transfers the CD behavior of the $\alpha $%
-peak to a constant, $y=0$, at its low frequency side ($x<0)$ and to a
linear decrease, $y=-x,$ (solid lines in Fig. \ref{scal}) at its
high-frequency side ($x>0$). The excess wing shows up as a more shallow
linear decrease deviating from the $y=-x$ line above $x\approx 5$. Due to
the large frequency region available for the present work this linear
decrease can be observed up to $x=20$ for PC while the maximum value
observed up to now was $x\approx 11$ \cite{Dendz}. Due to this larger range
it becomes obvious that the curves for glycerol and PC show significant
deviations from each other at $x\gtrsim 10$ as demonstrated in the upper
inset of Fig. \ref{scal}. In contrast, the original Nagel-scaling \cite
{Nagscal} scales the curves for both materials perfectly well in this
high-frequency region \cite{wing}. As mentioned above, the wing seems to
merge with the $\alpha $-peak above about $200$ K. In Fig. \ref{scal} this
corresponds to a direct transition from $y=-x$ behavior to the non-scaling
linear decrease corresponding to the minimum region (see lower inset of Fig. 
\ref{scal}). Here only two linear regions for $x>0$ can be distinguished
while at lower temperatures three such regions are seen (associated with $%
\alpha $-peak, excess wing, and minimum). For glycerol a similar merging of
wing and $\alpha $-peak is observed at temperatures above about $270$ K.

Finally we want to mention that it is also possible to describe the excess
wing by an additional relaxation process superimposed to the $\alpha $%
-process \cite{diel}. Such secondary processes, usually termed (slow) $\beta 
$-processes, can often be ascribed to an internal change of the molecular
conformation. However, in the light of the universal properties of the wing
for different glass-forming liquids revealed by the Nagel-scaling, it is
unreasonable to ascribe it to this type of $\beta $-relaxations which depend
on the specific molecular structure. The finding that secondary relaxation
processes can show up also in simple glass formers led to the assumption of
a more fundamental reason for these so-called Johari-Goldstein $\beta $%
-processes \cite{Johari}. Also recent theoretical developments \cite{cousin}
within the coupling model \cite{Ngai} may point in the direction of a
universal slow $\beta $-relaxation, closely connected to the $\alpha $%
-process. It cannot be excluded that the wing and these possible intrinsic $%
\beta $-relaxations are manifestations of the same microscopic mechanism.

\subsection{Fast dynamics}

\subsubsection{\protect\bigskip $\protect\varepsilon ^{\prime \prime }(%
\protect\nu )$-minimum}

\label{secdiscfastmin}In Fig. \ref{mctfits}, a magnified view of the
high-frequency region of $\varepsilon ^{\prime \prime }(\nu )$ is shown. In
the inset we tried to analyze the minimum region in terms of a simple
crossover from $\alpha $-relaxation or excess wing to the boson peak. For
the low frequency wing of the boson peak a linear or steeper increase of $%
\varepsilon ^{\prime \prime }(\nu )$ can be assumed as is commonly found for
a variety of glass-formers \cite{nsrev,Cum,Strom}. Indeed, for the lowest
temperatures investigated, where vibrational contributions can be assumed to
become dominant, a linear increase of $\varepsilon ^{\prime \prime }(\nu )$
(dashed line in Fig. \ref{mctfits}) is approached. Also, in order for the
boson peak to appear in the light- or neutron-scattering spectra (where it
was first observed), $\varepsilon ^{\prime \prime }\thicksim \nu S(q,\nu )$
must increase steeper than linearly towards the boson peak. The dashed lines
in the inset of Fig. \ref{mctfits} have been calculated by a sum of two
power laws, 
\begin{equation}
\varepsilon ^{\prime \prime }\thicksim c_{b}\nu ^{-b}+c_{n}\nu ^{n}
\label{pheno}
\end{equation}
with $n=1$ \cite{remadd}. $b$ was chosen to match the power law seen in $%
\varepsilon ^{\prime \prime }(\nu )$ between $1$ GHz and the $\varepsilon
^{\prime \prime }(\nu )-$minimum. Clearly there is no way to obtain a
reasonable fit of the experimental data in this way. This analysis clearly
implies the presence of additional fast processes in this region, usually
termed fast $\beta $-processes. Similar findings were obtained for a variety
of glass-formers \cite{Cum,Lunkigly,Lunkiorl,Brand}.

Fast processes are not considered in the EFV model \cite{FV,EFV}, which
provides the best fits to the dynamics of the $\alpha $-process (section \ref
{discalp}). Within this model, molecules are assumed to move within cages
formed by their next neighbors and diffusional motion occurs by hopping into
''free volume'' available to the molecule. It is interesting that this
picture is qualitatively similar to that of MCT where fast processes are a
natural consequence of the theory. But a theoretical investigation of these
possible fast processes in the EFV is missing. In the FLD model \cite{Kiv}
the $\alpha $-relaxation is identified with the restructuring of the FLDs
and occurs on a length scale given by their characteristic size. Within the
FLD framework there is a second length scale, the correlation length $\xi $
of the LPS. It was argued \cite{Kiv} that the experimentally observed fast $%
\beta $-processes may be ascribed to fast relaxations taking place on this
smaller length scale of the FLD model. However, up to now a theoretical
elaboration of the fast processes within FLD theory is missing.

Another model predicting fast processes is the coupling model (CM)\ of Ngai
and coworkers \cite{Ngai}. Here the fast process is part of the $\alpha $%
-relaxation, defined in the time domain. The $\alpha $-relaxation is assumed
to be composed of a fast exponential decay with relaxation time $\tau _{f}$
at short times $t<t_{c}$, and a slower KWW behavior with relaxation time $%
\tau _{s}$ and stretching parameter $\beta _{s}$ at $t>t_{c}$. The crossover
time $t_{c}$ was stated to be of the order of ps for most systems \cite{Ngai}%
. Demanding continuity at $t_{c}$ leads to the relation: 
\begin{equation}
\tau _{f}=\tau _{s}^{\beta _{s}}t_{c}^{1-\beta _{s}}  \label{eqCM}
\end{equation}

This allows for the determination of the relaxation time $\tau _{f}$ of the
fast process, if the KWW-parameters of the slow process are known. The
parameters determined from the KWW-fits of the $\alpha $-relaxation in PC
(section \ref{discalp}) result in a factor $\tau _{s}/\tau _{f}\approx 1.3$
for the highest temperature and $\tau _{s}/\tau _{f}\approx 10^{4}$ for $153$
K. Obviously, the fast process inherent in the CM is located at much lower
frequencies than the $\varepsilon ^{\prime \prime }(\nu )$-minimum and
therefore cannot be brought into play for the explanation of the excess
intensity in this region. Instead additional contributions need to be
assumed but, as shown in the beginning of this section, a combination of $%
\alpha $-process (even including the CM%
%TCIMACRO{\UNICODE{0xb4}}%
%BeginExpansion
\'{}%
%EndExpansion
s fast process) and boson peak alone cannot take account of the experimental
data. Further contributions may arise \cite{Ngaipriv}, e.g., from a constant
loss term as early proposed by Wong and Angell \cite{Wong} and/or a possible 
$\varepsilon ^{\prime \prime }\thicksim \nu ^{0.3}$ behavior, both seemingly
universal features in the dielectric response of glassy ionic conductors 
\cite{colo,coloexpl}. Possible microscopic explanations for these
contributions were proposed in \cite{coloexpl}. Indeed, the data in PC near
the $\varepsilon ^{\prime \prime }(\nu )$-minimum can well be described by
adding a term $\varepsilon _{c}+c_{3}\nu ^{0.3}$ to the ansatz of Eq. \ref
{pheno} \cite{fukglas}. In this way, by adjusting the parameters to achieve
a smooth transition to the CD-curves describing the $\alpha $-peak, quite
impressive fits over 17 decades of frequency are possible. Of course this
could be expected considering the large amount of parameters involved in the
fitting, but nevertheless it cannot be excluded that such a simple
superposition ansatz of different independent contributions is correct.

Very recently a model was proposed considering a low-frequency
relaxation-like part of the vibration susceptibility function responsible
for the boson peak \cite{Novi}. This relaxation-like response was shown to
arise from anharmonicity of vibrations and invoked to explain the
quasielastic line seen in light and neutron scattering. This quasielastic
contribution corresponds to the fast $\beta $-process seen in the region
below the boson peak in the susceptibility spectra. Within this framework,
the fast process should show up as peak with $\varepsilon ^{\prime \prime
}\thicksim \nu ^{\alpha }$ and $\varepsilon ^{\prime \prime }\thicksim \nu
^{-1}$ at its low and high-frequency side, respectively. For the absolute
value of the exponent $\alpha $,\ reasonable values ranging between 0.375
and 1 were given in \cite{Novi}, depending on the system. Of course no peak
is seen between $\alpha $- and boson peak in the spectra of PC (Figs. \ref
{eps2} and \ref{mctfits}) but the $\nu ^{-1}$-wing of the fast $\beta $%
-process may be obscured by the domination of the boson peak at higher
frequencies. A severe argument against this picture is the smooth transition
of $\varepsilon ^{\prime \prime }(\nu )$ from the minimum to the boson peak
with only one power law, which seems unlikely to result from the
superposition of the $\nu ^{\alpha }$-wing of the fast $\beta $-process and
the low frequency wing of the boson peak (steeper than $\nu ^{1}$, see
above). However, this possibility cannot fully be excluded and at least this
model provides an explanation for the sublinear increase seen for $\nu >\nu
_{\min }$. It is interesting, that within this theoretical framework a
crossover temperature of the fast relaxation was obtained which was found to
be of similar magnitude as the critical temperature $T_{c}$ of MCT \cite
{Novi}.

As mentioned in the introduction, a fast process in the minimum region of $%
\varepsilon ^{\prime \prime }(\nu )$ is one of the main outcomes of MCT, a
prediction which led to the huge interest in this high-frequency region.
Within idealized MCT, above $T_{c}$, the high and low-frequency wings close
to the minimum of $\varepsilon ^{\prime \prime }(\nu )$ are described by
power laws with exponents $a$ and $b$, respectively. The minimum region can
be approximated by the interpolation formula \cite{mct,mctrev}:

\begin{equation}
\varepsilon ^{\prime \prime }(\nu )=\varepsilon _{\min }[a(\nu /\nu _{\min
})^{-b}+b(\nu /\nu _{\min })^{a}]/(a+b)  \label{eqMCTmin}
\end{equation}
$\nu _{min}$ and $\varepsilon _{min}^{\prime \prime }$ denote position and
amplitude of the minimum, respectively. The exponents $a$ and $b$ are
temperature independent and constrained by the exponent parameter $\lambda
=\Gamma ^{2}(1-a)/\Gamma (1-2a)=\Gamma ^{2}(1+b)/\Gamma (1+2b)$ where $%
\Gamma $ denotes the Gamma function. This restricts the exponent $a$ to
values below $0.4$, i.e. a significantly sublinear increase of $\varepsilon
^{\prime \prime }(\nu )$ at frequencies above the minimum is predicted. We
found a consistent description of the $\varepsilon ^{\prime \prime }$-minima
at $T\geq 193$ K using $\lambda =0.78$ which implies $a=0.29$ and $b=0.5$
(solid lines in Fig. \ref{mctfits}). The obtained $\lambda $ agrees with
that determined from the preliminary data published previously \cite
{Lunkiorl,Lunkikyo,Lunkibos}. It is identical to $\lambda $ deduced from
light scattering results \cite{Du} and also consistent with a recent
analysis of solvation dynamics experiments on PC \cite{Berg}. In addition, a
preliminary analysis of the susceptibility minimum from recent neutron
scattering measurements \cite{Ohl} yields $\lambda =0.75$, consistent with
the present value. At high frequencies the fits with Eq. \ref{eqMCTmin} are
limited by the onset of the decrease at the high-frequency wing of the boson
peak. Quite a different behavior was seen in glycerol \cite{Lunkibos,fukglas}
where the MCT fits are limited at high frequencies by an additional
increase, that cannot be described by Eq.\ \ref{eqMCTmin}. In \cite
{Lunkibos,fukglas} it was argued that this deviation in glycerol is due to
the vibrational boson peak contribution which is not taken into account by
Eq. \ref{eqMCTmin}. In PC this contribution seems to be of less importance.
This is in accord with the finding of Sokolov {\it et al}. \cite{Sok} that
the amplitude ratio of boson peak and fast process is largest for strong
glass formers, glycerol being much stronger than PC (see also section \ref
{secdiscfastbos}). The deviations of data and fits, seen at low frequencies
in Fig. \ref{mctfits}, can be ascribed to the growing importance of the $%
\alpha $-relaxation which is not taken into account by the simple
interpolation formula, Eq. \ref{eqMCTmin}. The critical temperature $T_{c}$
should manifest itself in the temperature dependence of the $\varepsilon
^{\prime \prime }(\nu )$-minimum. For $T>T_{c}$ MCT predicts the following
relations: $\nu _{min}\thicksim (T-T_{c})^{1/(2a)}$ and $\varepsilon
_{min}^{\prime \prime }\thicksim (T-T_{c})^{%
%TCIMACRO{\UNICODE[m]{0xbd}}%
%BeginExpansion
{\frac12}%
%EndExpansion
}$. Figure \ref{mopa} shows the results for $\varepsilon _{min}^{\prime
\prime }(T)$, $\nu _{min}(T)$, and also for $\nu _{\tau }(T)$ (same data as
in Fig. \ref{tau}). Here representations have been chosen that lead to
straight lines that extrapolate to $T_{c}$ if the predicted critical
temperature dependences are obeyed. Indeed, as indicated by the solid lines,
all three data sets can be described consistently with a critical
temperature $T_{c}\approx 187$ K, in agreement with the results determined
from the preliminary data published previously \cite
{Lunkiorl,Lunkikyo,Lunkibos}. This lies in the same range as the $T_{c}$
obtained from light scattering \cite{Du} ($187$ K), neutron scattering \cite
{Borj} ($180$ - $200$ K), and solvation dynamics experiments \cite{Berg} ($%
176$ K). Also the $T_{c}\approx 186$ K, deduced from a preliminary analysis
of the susceptibility minimum from recent neutron scattering experiments 
\cite{Ohl}, agrees well with the present value. For temperatures near $T_{c}$
the $\nu _{\tau }$-data deviate from the predicted behavior as also seen in
Fig. \ref{tau}(c). Within MCT this can be ascribed to a smearing out of the
critical behavior near $T_{c}$ due to hopping processes which are considered
in extended versions of MCT only \cite{mctrev,mctext}. In addition, the
above critical temperature dependences of $\varepsilon _{min}^{\prime \prime
}$, $\nu _{min}$ and $\nu _{p}$ should be valid only for temperatures not
too far above $T_{c}$ and the deviations seen for $\nu _{min}$ and $%
\varepsilon _{min}^{\prime \prime }$ at high temperatures (Fig. \ref{mopa})
may be due to this fact. Therefore there is a problem concerning the proper
choice of the temperature range to be used for the determination of $T_{c}$.
While the critical behavior proposed in Fig. \ref{mopa} seems reasonable,
some uncertainties concerning the value of $T_{c}$ remain which can only be
clarified by an analysis within the extended MCT.

MCT predicts a significant change in the behavior of $\varepsilon ^{\prime
\prime }(\nu )$ at $T_{c}$: For $T<T_{c}$, within idealized MCT $\varepsilon
^{\prime \prime }(\nu )$ should exhibit a so-called ''knee'' at a frequency $%
\nu _{k}$, i.e. a change of power law from $\varepsilon ^{\prime \prime
}\sim \nu ^{a}$ at $\nu >$ $\nu _{k}$ to $\varepsilon ^{\prime \prime }\sim
\nu $ at $\nu <\nu _{k}$. Within idealized MCT, at $T<T_{c}$ the minimum and
the $\alpha $-peak should vanish but both phenomena are restored by the
hopping processes invoked in extended MCT \cite{mctrev,mctext}. At first,
from Fig. \ref{mctfits} it becomes obvious that the fits with eq. \ref
{eqMCTmin} with $\lambda =0.78$ will no\ longer work at $T<T_{c}$ because
the high frequency wing of the minimum becomes successively steeper below
this temperature, i.e. the frequency dependence of $\varepsilon ^{\prime
\prime }(\nu )$ changes significantly below $T_{c}$. In Fig. \ref{mctfits},
a $\varepsilon ^{\prime \prime }\thicksim \nu ^{a}$ behavior with $a=0.29$
as determined for $T>T_{c}$ is shown as dashed line matching the region just
below the boson peak for the $183$ K and $173$ K curves. For lower
frequencies the experimental data exhibit a downward deviation from the $\nu
^{a}$-extrapolations, i.e. really a knee appears. Whatsoever, in light of
the rather large experimental uncertainties in this region it seems quite
audacious to assign the observed feature to the knee of MCT. According to
MCT, $\nu _{k}$ and the amplitude at the knee, $\varepsilon _{k}^{\prime
\prime }$, should also exhibit critical behavior, $\nu _{k}\thicksim
(T_{c}-T)^{1/(2a)}$ and $\varepsilon _{k}^{\prime \prime }\thicksim
(T_{c}-T)^{%
%TCIMACRO{\UNICODE[m]{0xbd}}%
%BeginExpansion
{\frac12}%
%EndExpansion
}$ with the proportionality factors in close connection of the
proportionality factors in the critical behavior of the minimum \cite
{Gotzepriv}. In Fig. \ref{mopa}(a) and (b) the two values read off from Fig. 
\ref{mctfits} are denoted as pluses. Having in mind the extreme
uncertainties in the knee position, the data seem to be consistent with a $%
T_{c}$ of $187$ K. However, the position of the knee is not in accord with
the position of the miniumum above $T_{c}$ \cite{Gotzepriv}. Overall, a lack
of significance of the observed knee-feature has to be stated but at least
the data show where the knee may be located and where more experimental work
has to be done to check for its presence and temperature evolution in PC.

\subsubsection{Comparison with scattering results}

The comparison of dielectric and the scattering results is shown in Fig. \ref
{vgl}. As mentioned above (section \ref{resfast}) the significantly larger
ratio of $\alpha $- and boson peak amplitude for dielectric compared with
light scattering results seems to be a rather universal property of
glass-forming materials \cite{Lunkibos,Schn,Wahn,Schsim}. Recently, an
explanation for this finding was given by considering the different
dependencies of the probes on orientational fluctuations \cite{Leb}. In
addition, in a recent theoretical work \cite{Schkug} MCT was generalized to
molecular liquids of molecules with orientational degrees of freedom. The
resulting MCT equations were solved for a system of dipolar hard spheres and
a larger ratio of $\alpha $- and boson peak amplitude for $\varepsilon
^{\prime \prime }$ was obtained in accordance with the experimental
findings. Further, it is one of the main predictions of MCT \cite
{mct,mctrev,mctext} that the same parameters $T_{c\text{ }}$and $\lambda $\
should arise from all observables coupling to the density fluctuations. This
is indeed the case for propylene-carbonate (see above). But also the
position of the $\varepsilon ^{\prime \prime }$- (respectively $\chi
^{\prime \prime }$-) minima should be the same, independent of the
experimental method which is also valid when including orientational degrees
of freedom \cite{Schpriv}. However, as seen in Fig. \ref{vgl}, this seems
not to be the case. An explanation for a similar behavior found from
molecular dynamics simulations of a system of rigid diatomic molecules was
given in terms of $180%
%TCIMACRO{\UNICODE[m]{0xb0}}%
%BeginExpansion
{{}^\circ}%
%EndExpansion
$ flips of the non-spherical molecules which cannot be taken into account by
a hard sphere model \cite{Schsim}. In this context, it may be mentioned that
for the mobile-ion glass-former CKN a rather good agreement of the minimum
position from different experimental methods was found \cite{Lunkikyo}. Here
orientational degrees of freedom are not important which corroborates the
above explanation in terms of $180%
%TCIMACRO{\UNICODE[m]{0xb0}}%
%BeginExpansion
{{}^\circ}%
%EndExpansion
$ flips.

\subsubsection{Boson peak}

\label{secdiscfastbos}The boson peak is a universal feature of glass-forming
materials showing up in light and neutron scattering but also as excess
contribution in specific heat measurements. A variety of explanations of the
boson peak has been proposed, e.g., in terms of the soft potential model 
\cite{soft}, phonon localization models \cite{phonloc}, or a model of
coupled harmonic oscillators with a distribution of force constants \cite
{Schirm}. Also MCT includes contributions leading to a peak at THz as was
shown using a schematic two-correlator model \cite{Fra}. However, up to now
there is no consensus concerning the microscopic origin of this phenomenon.
It cannot be the aim of the present paper to compare the results with the
proposed models, especially in the light of the rather restricted data base
for $\nu >1$ THz. However, some experimental facts deserve to be emphasized
which may be of importance for the understanding of the boson peak. Figure 
\ref{pcgly} shows the high frequency region of $\varepsilon ^{\prime \prime
}(\nu )$ for PC and glycerol. Clearly, both materials exhibit quite
different behavior in the boson peak region: For PC only one power law is
seen at frequencies $\nu >\nu _{\min }$ forming simultaneously the high
frequency wing of the $\varepsilon ^{\prime \prime }$-minimum and the low
frequency wing of the boson peak. In marked contrast, for glycerol two
regimes can be distinguished: Directly above $\nu _{\min }$, there is a
rather shallow increase of $\varepsilon ^{\prime \prime }(\nu )$\ which
together with the left wing of the minimum can be well fitted with the MCT
prediction, Eq. \ref{eqMCTmin} \cite{Lunkigly,Lunkibos,fukglas}. At higher
frequencies a very steep increase appears, approaching $\varepsilon ^{\prime
\prime }\thicksim \nu ^{3}$ for $T\rightarrow T_{g}$. This different
behavior of PC and glycerol also becomes obvious in the scaling plot of Fig. 
\ref{scal}. For both materials the $\varepsilon ^{\prime \prime }(\nu )$
-minimum corresponds to the deviation from the master curve, that shows up
as a third linear region in the scaling plot. But for glycerol in addition a
minimum is seen in Fig. \ref{scal}. In the scaling plot this minimum marks
the transition point between the two regimes at $\nu >\nu _{\min }$,
described above. It seems that in glycerol the fast $\beta $-process giving
rise to the shallow minimum in $\varepsilon ^{\prime \prime }(\nu )$ or the
third linear region in the scaling plot, is obscured at higher frequencies
by the boson peak contribution. This arguing seems to be in accord with the
finding of Sokolov {\it et al}. \cite{Sok} of a higher amplitude-ratio of
boson peak and fast process for strong glass formers. However, at least near 
$T_{g}$ this ratio, e.g. measured by comparing boson-peak and minimum
amplitude in Fig. \ref{pcgly}, is quite similar for both materials. For high
temperatures, $T\geq 193$ K for PC and $T\geq 253$ K for glycerol, the fast $%
\beta $-process in PC has indeed a higher amplitude if curves with similar
values of $\nu _{p}$ are compared. This fact also explains the good fits
with MCT up to the boson peak frequency (section \ref{secdiscfastmin}) in
PC. However, for lower temperatures the different characteristics of PC and
glycerol has to be ascribed either to a more shallow left wing of the boson
peak for PC ($\varepsilon ^{\prime \prime }\thicksim \nu ^{1}$ instead of $%
\varepsilon ^{\prime \prime }\thicksim \nu ^{3}$) or to a steeper increase
of the high frequency wing of the minimum, thereby obscuring the steeper
boson peak wing. Interestingly, while there might be some problems for the
new model by Novikov \cite{Novi} in explaining the PC results (see section 
\ref{secdiscfastmin}), glycerol seems to fit into this picture: Below the
boson peak, two distinct regions are seen which can be ascribed to the
relaxation-like and the main part of the vibrational excitations. Concerning
MCT, it has to be mentioned that the light scattering results on glycerol 
\cite{Wutt} exhibit quite similar characteristics as the dielectric results 
\cite{Schn}. The light scattering results were successfully described using
the schematic two-correlator MCT model mentioned above \cite{Fra}. But it is
not clear if MCT is able to provide an explanation for the qualitative
difference of the glycerol and PC dielectric data.

\subsection{CONCLUSIONS}

Dielectric data on glass-forming PC in an exceptionally broad frequency
range have been presented and compared with the results from light
scattering and to broadband dielectric results on glycerol. A comparison of
the findings concerning $\alpha $-relaxation, excess wing, $\varepsilon
^{\prime \prime }(\nu )$-minimum, and boson peak on various theoretical and
phenomenological predictions has been performed. The temperature dependence
of the $\alpha $-relaxation time is in accord with the EFV and the FLD
models and at high temperatures also with that of idealized MCT. The
temperature dependences of the relaxation strength and the width parameter
of the $\alpha $-relaxation were determined and some differences to the
findings from previous works were found. The excess wing in PC was shown to
follow the modified Nagel scaling proposed by Dendzik et al. \cite{Dendz}
but at high frequencies the curves of PC and glycerol do not scale onto each
other. Overall, it is not possible to arrive at a decision in favor or
against one of the models on the dynamics of glass-forming materials from
the results obtained at the first 15 decades of frequency alone.

At higher frequencies above about 1 GHz a minimum in $\varepsilon ^{\prime
\prime }(\nu )$ is detected. In this region, clearly fast processes
contribute to the dielectric loss, a finding for which most models provide
no explanation. While the minimum can be well described by the sum of power
laws and a constant loss contribution \cite{fukglas}, the theoretical
foundation for these contributions is unclear at present but currently is
being developed. In contrast, MCT provides a microscopic explanation for the
high-frequency data, at least at high temperatures. In addition, the
analysis within MCT yields values for $T_{c}$ and $\lambda $, that are in
agreement with those determined from other experimental methods. Also the
differences in the susceptibilities obtained from dielectric and light
scattering experiments seem to be in accord with recent extensions of MCT.
At low temperatures, $T<$ $T_{c}$, at least a qualitative agreement of the
results (presence of a fast process and possible occurrence of a ''knee'')
with MCT can be stated, but more work is needed for a quantitative
comparison.

Overall, certain aspects of the results found in the present work can be
explained by different models, but in our judgement MCT provides the most
consistent picture providing explanations for the largest variety of
experimental facts, including the high-frequency processes. This may be
partly due to the large amount of predictions made by MCT, covering almost
all aspects of the dynamic response of glass-forming materials, while other
models are somewhat restricted in this respect. However, just this agreement
of the data with the many different predictions of MCT speaks in favor of
the theory. Admittedly, it is not possible to arrive at a final decisive
conclusion concerning the applicability of MCT for PC and alternative
explanations may be possible. Also many questions need to be solved, e.g.
the behavior below $T_{c}$ should be compared with the extended MCT, the
differences to the scattering results should be compared quantitatively to
MCT (considering orientational degrees of freedom of dipolar, non-spherical
molecules), and the qualitatively different behavior of PC and glycerol in
the boson peak region should be addressed. Finally, in the boson peak region
we find significant differences between glycerol and PC which wait for a
theoretical explanation. But also more experimental work is needed, e.g.
concerning the temperature evolution of the boson peak and a higher
precision of the data at low temperatures in order to investigate in detail
the possible occurrence of a knee in $\varepsilon ^{\prime \prime }(\nu )$
of PC.

\acknowledgments

We gratefully acknowledge stimulating discussion with C.A. Angell, R.
B\"{o}hmer, H.Z. Cummins, W. G\"{o}tze, K.L. Ngai, M. Ohl, R. Schilling, W.
Schirmacher, and J. Wuttke. We thank M. Ohl and J. Wuttke for providing the
neutron scattering results of ref. \cite{Ohl} prior to publication. This
work was supported by the Deutsche Forschungsgemeinschaft, Grant-No.
LO264/8-1 and partly by the BMBF, contract-No. 13N6917.

\begin{figure}[tbp]
\caption{Frequency dependence of the dielectric constant in PC at various
temperatures. The solid lines are fits with the CD function performed
simultaneously on $\protect\varepsilon ^{\prime \prime }$. The dotted line
is a fit with the Fourier transform of the KWW function.}
\label{eps1}
\end{figure}

\begin{figure}[tbp]
\caption{Frequency dependence of the dielectric loss in propylene carbonate
at various temperatures. The solid lines are fits with the CD function, the
dotted line is a fit with the Fourier transform of the KWW law, both
performed simultaneously on $\protect\varepsilon ^{\prime }$. The
dash-dotted line indicates a linear increase. The FIR results have been
connected by a dashed line to guide the eye. The inset shows $\protect\nu _{%
\protect\tau }=1/(2\protect\pi <\protect\tau >)$ as resulting from the CD
(circles) and KWW fits (pluses) in an Arrhenius representation. The line is
a fit using the VFT expression, Eq. \ref{eqVFT} with $T_{VF}=132$ K, $D=6.6$%
, and $\protect\nu _{0}=3.2\times 10^{12}$ Hz. }
\label{eps2}
\end{figure}

\begin{figure}[tbp]
\caption{Relaxation strength $\Delta \protect\varepsilon $ (a) and width
parameter $\protect\beta $ (b) as obtained from simultaneous fits of $%
\protect\varepsilon ^{\prime }(\protect\nu )$ and $\protect\varepsilon %
^{\prime \prime }(\protect\nu )$\ using the CD function and the Fourier
transform of the KWW function. The solid line in (a) has been calculated
using the MCT prediction (see text). The solid lines in (b) indicate a
linear increase.}
\label{betchi}
\end{figure}

\begin{figure}[tbp]
\caption{Frequency dependence of the dielectric loss in propylene carbonate
compared to the susceptibility as calculated from the light scattering
results (data digitized from \protect\cite{Du}). The light scattering data
sets have been vertically shifted to give a comparable intensity of the
boson peak.}
\label{vgl}
\end{figure}

\begin{figure}[tbp]
\caption{Temperature dependence of $\protect\nu _{\protect\tau }=1/(2\protect%
\pi <\protect\tau >_{CD})$ as determined from simultaneous CD fits of $%
\protect\varepsilon ^{\prime }(\protect\nu )$ and $\protect\varepsilon %
^{\prime \prime }(\protect\nu )$\ in Arrhenius representation. (a) solid
line: fit with VFT behavior (Eq. \ref{eqVFT}, $\protect\nu _{0}=3.2\times
10^{12}$ Hz, $D=6.6$, $T_{VF}=132$ K); dashed line: Arrhenius behavior ($%
\protect\nu _{0}=9.0\times 10^{12}$ Hz, $E=2290$ K,). (b) solid line: fit
with EFV theory (Eq. \ref{eqEFV}, $A=10.7$, $B=309$ K, $C=4.82$ K, $T_{0}=162
$ K). (c) solid line: fit with MCT (Eq. \ref{eqMCTfp}, $T_{c}=187$ K, $%
\protect\gamma =2.72$, $c=12900$). $T_{c}$ and $\protect\gamma $ were fixed
at the values determined from the evaluation of the $\protect\varepsilon %
^{\prime \prime }(\protect\nu )$-minimum. Dashed line: fit with VFT behavior
for $T<200$ K (Eq. \ref{eqVFT}, $\protect\nu _{0}=4.9\times 10^{13}$ Hz, $%
D=8.4$, $T_{VF}=128$ K). (d) solid line: fit with FLD theory (Eq. \ref{Kiveq}%
, $T^{\ast }=215$ K, $\protect\nu _{\infty }=2.03\times 10^{14}$ Hz, $%
E_{\infty }=3090$ K, $F=457$); dashed line: high-temperature asymptotic
Arrhenius behavior.}
\label{tau}
\end{figure}

\begin{figure}[tbp]
\caption{Scaling plot of $\protect\epsilon ^{\prime \prime }(\protect\nu )$
for PC and glycerol according to \protect\cite{Dendz}. Note that for clarity
reasons both data-sets are shifted with respect to each other. The lines
show the asymptotic behavior for a CD function at $\protect\nu <\protect\nu
_{p}$ and $\protect\nu >\protect\nu _{p}$. The lower\ inset is a magnified
plot for the transition region between CD and excess wing contribution in
PC. Here the 295 K points are connected by a solid line to guide the eye.
The upper inset shows one glycerol and one PC curve in the wing region.}
\label{scal}
\end{figure}

\begin{figure}[tbp]
\caption{$\protect\varepsilon ^{\prime \prime }(\protect\nu )$\ at high
frequencies for various temperatures. The solid lines are fits with MCT, Eq. 
\ref{eqMCTmin}, with $a=0.29$ and $b=0.5$ ($\protect\lambda =0.78$). The
dashed lines demonstrate $\protect\nu ^{a}$ behavior for $T<T_{c}$ and may
indicate the occurrence of a ''knee''. The dash-dotted line indicates linear
behavior. The dotted line is drawn to guide the eye. The dashed lines in the
inset have been calculated with Eq. \ref{pheno}. }
\label{mctfits}
\end{figure}

\begin{figure}[tbp]
\caption{Temperature dependence of the amplitude (a) and position (b) of the 
$\protect\varepsilon ^{\prime \prime }(\protect\nu )$-minimum and of the $%
\protect\alpha $-peak position $\protect\nu _{\protect\tau }=1/(2\protect\pi
<\protect\tau >_{CD})$ (c) of PC. $\protect\varepsilon _{\min }^{\prime
\prime }$ and $\protect\nu _{\min }$ have been taken from the fits with Eq. 
\ref{eqMCTmin}. Representations have been chosen that result in straight
lines according to the predictions of the MCT. The solid lines extrapolate
to a $T_{c}$ of 187 K for all three quantities as indicated by the arrows.
The pluses give an estimate for the position of the tentative ''knee'', that
may be suspected from Fig. \ref{mctfits}. }
\label{mopa}
\end{figure}

\begin{figure}[tbp]
\caption{$\protect\varepsilon ^{\prime \prime }(\protect\nu )$ in the
high-frequency region for PC (a) and glycerol (b). The dash-dotted lines
indicate power laws as noted in the figure. The dashed line in (a) was drawn
to guide the eye.}
\label{pcgly}
\end{figure}

%\end{multicols}


\begin{references}
\bibitem{hist}  see, e.g., M. Cable, in: {\it Glasses and Amorphous Materials%
}, ed. J. Zarzycki (Materials Science and Technology, Vol. 9, VCH, Weinheim,
1991), p. 1.

\bibitem{mct}  U. Bengtzelius, W. G\"{o}tze, and A. Sj\"{o}lander, J. Phys.
C {\bf 17}, 5915 (1984); E. Leutheuser, Phys. Rev. A {\bf 29}, 2765 (1984);
W. G\"{o}tze, Z. Phys. B {\bf 60}, 195 (1985).

\bibitem{mctrev}  for a review of MCT, see: W. G\"{o}tze and L. Sj\"{o}gren,
Rep. Progr. Phys {\bf 55}, 241 (1992).

\bibitem{mctext}  W. G\"{o}tze and L. Sj\"{o}gren, Z. Phys. B {\bf 65}, 415
(1987).

\bibitem{Ngai}  K.L. Ngai, Comments Solid State Phys. {\bf 9}, 121 (1979);
K.L. Ngai, C.H. Wang, G. Fytas, D.L. Plazek, and D.J. Plazek, J. Chem. Phys. 
{\bf 86}, 4768 (1987).

\bibitem{Chamb}  R.V. Chamberlin, Phys. Rev B {\bf 48}, 15638 (1993).

\bibitem{Kiv}  D. Kivelson, S.A. Kivelson, X-L. Zhao, Z. Nussinov, and G.
Tarjus, Physica A {\bf 219}, 27 (1995); G. Tarjus, D. Kivelson, and S.
Kivelson, in: {\it Supercooled Liquids: Advances and Novel Applications},
ed. J.T. Fourkas, D. Kivelson, U. Mohanty, and K.A. Nelson (ACS
Publications, Washington, DC, 1997) p. 67.

\bibitem{Nagscal}  P.K. Dixon, L. Wu, S.R. Nagel, B.D. Williams, and J.P.
Carini, Phys. Rev. Lett. {\bf 65}, 1108 (1990).

\bibitem{Nagsus}  N. Menon and S.R. Nagel, Phys. Rev. Lett {\bf 74}, 1230
(1995).

\bibitem{Rich}  see collection of papers in: {\it Disorder Effects on
Relaxational Processes}, ed. R. Richert and A. Blumen (Springer, Berlin,
1994).

\bibitem{nsrev}  for a review of neutron scattering results, see: W. Petry
and J. Wuttke, Transp. Theory Statist. Phys. {\bf 24}, 1075 (1995)

\bibitem{Cum}  H.Z. Cummins, G. Li, Y.H. Hwang, G.Q. Shen, W.M. Du, J.
Hernendez, and N.J. Tao, Z. Phys. B {\bf 103}, 501 (1997).{\bf \ }

\bibitem{procs}  see collection of papers in: {\it Dynamics of Glass
Transition and Related Topics}, ed. T. Odagaki, Y Hiwatari, and J. Matsui
[Progr. Theor. Phys. Suppl. {\bf 126} (1997)]; {\it Structure and Dynamics
of Glasses and Glass Formers}, ed. C.A. Angell, K.L. Ngai, J. Kieffer, T.
Egami, and G.U. Nienhaus (Materials Research Society symposium proceedings,
Vol. 455, Pittsburgh, 1997).

\bibitem{Wutt}  J. Wuttke, J. Hernandez, G. Li, G. Coddens, H.Z. Cummins, F.
Fujara, W. Petry, and H. Sillescu, Phys. Rev. Lett. {\bf 72}, 3052 (1994);
J. Wuttke, W. Petry, G. Coddens, and F. Fujara, Phys. Rev. E {\bf 52}, 4026
(1995).

\bibitem{Du}  W.M. Du, G. Li, H.Z. Cummins, M. Fuchs, J. Toulouse, and L.A.
Knauss, Phys. Rev. {\bf E} 49, 2192 (1994).

\bibitem{PimPRE}  A. Pimenov, P. Lunkenheimer, H. Rall, R. Kohlhaas, A.
Loidl, and R. B\"{o}hmer, Phys. Rev. E {\bf 54,} 676 (1996).

\bibitem{Funke}  K.L. Ngai, C. Cramer, T. Saatkamp, and K. Funke, in: {\it %
Proceedings of the Workshop on Non-Equilibrium Phenomena in Supercooled
Fluids, Glasses, and Amorphous Materials}, eds. M. Giordano, D. Leporini,
and M. Tosi (World Scientific, Singapore, 1996) p. 3.

\bibitem{Lunkigly}  P. Lunkenheimer, A. Pimenov, M. Dressel, Yu. G.
Goncharov, R. B\"{o}hmer, and A. Loidl, Phys. Rev. Lett. {\bf 77}, 318
(1996).

\bibitem{Lunkiorl}  P. Lunkenheimer, A. Pimenov, M. Dressel, B. Gorshunov,
U. Schneider, B. Schiener, and A. Loidl, in: {\it Supercooled Liquids:
Advances and Novel Applications}, ed. J.T. Fourkas, D. Kivelson, U. Mohanty,
and K.A. Nelson (ACS Publications, Washington, DC, 1997) p. 168.

\bibitem{Lunkickn}  P. Lunkenheimer, A. Pimenov, and A. Loidl, Phys. Rev.
Lett. {\bf 78}, 2995 (1997).

\bibitem{diel}  A. Hofmann, F. Kremer, E.W. Fischer, and A. Sch\"{o}nhals,
in: {\it Disorder Effects on Relaxational Processes}, ed. R. Richert and A.
Blumen (Springer, Berlin, 1994), p. 309.

\bibitem{wing}  R. Brand, P. Lunkenheimer, U. Schneider, and A. Loidl,
submitted to Phys. Rev. Lett.

\bibitem{Wong}  J. Wong and C.A. Angell in {\it Glass: Structure by
Spectroscopy }(M. Dekker Inc., New York, Basel, 1974), p. 750.

\bibitem{Lunkikyo}  P. Lunkenheimer, A. Pimenov, M. Dressel, B. Schiener, U.
Schneider, and A. Loidl, Progr. Theor. Phys. Suppl. {\bf 126}, 123 (1997).

\bibitem{Lunkibos}  P. Lunkenheimer, A. Pimenov, M. Dressel, B. Gorshunov,
U. Schneider, B. Schiener, R. B\"{o}hmer, and A. Loidl, in: {\it Structure
and Dynamics of Glasses and Glass Formers}, ed. C.A. Angell, K.L. Ngai, J.
Kieffer, T. Egami, and G.U. Nienhaus (Materials Research Society symposium
proceedings, Vol. 455, Pittsburgh, 1997) p. 47.

\bibitem{Schn}  U. Schneider, P. Lunkenheimer, R. Brand, and A. Loidl, J.
Non-Cryst. Solids {\bf 235-237,} 173 (1998).

\bibitem{strong}  C.A. Angell, in {\it Relaxations in Complex Systems}, ed.
K.L Ngai and G.B. Wright (NRL, Washington, D.C., 1985), p.3.

\bibitem{m}  D.J. Plazek and K.L. Ngai, Macromolecules {\bf 24}, 1222
(1991); R. B\"{o}hmer and C.A. Angell, Phys. Rev. B{\bf \ 45}, 10091 (1992).

\bibitem{4glass}  A. Pimenov, P. Lunkenheimer, M. Nicklas, R. B\"{o}hmer, A.
Loidl, and C.A. Angell, J. Non-Cryst. Solids {\bf 220}, 93 (1997).

\bibitem{rol}  R. B\"{o}hmer, K.L. Ngai, C.A. Angell, and D.J. Plazek, J.
Chem. Phys. {\bf 99}, 4201 (1993).

\bibitem{Sok}  A.P. Sokolov, E. R\"{o}ssler, A. Kisliuk, and D. Quitmann,
Phys. Rev. Lett. {\bf 71}, 2062 (1993).

\bibitem{Mops}  F.I. Mopsik, Rev. Sci. Instrum. {\bf 55}, 79 (1984).

\bibitem{jappl}  R. B\"{o}hmer, M. Maglione, P. Lunkenheimer, and A. Loidl,
J. Appl. Phys. {\bf 65}, 901 (1989).

\bibitem{Volkov}  A.A. Volkov, Yu.G. Goncharov, G.V. Kozlov, S.P. Lebedev,
and A.M. Prokhorov, Infrared Phys. {\bf 25}, 369 (1985); A.A. Volkov{\it , }%
G.V. Kozlov, S.P. Lebedev, and A.M. Prokhorov, Infrared Phys. {\bf 29}, 747
(1989).

\bibitem{Born}  M. Born and E. Wolf, {\it Principles of Optics }(Pergamon
Press, Oxford, 1980).

\bibitem{AngPC}  C.A. Angell, L. Boehm, M. Oguni, and D.L. Smith, J.
Molecular Liquids {\bf 56}, 275 (1993).

\bibitem{Johari}  G.P. Johari and M. Goldstein, J. Chem. Phys. {\bf 53},
2372 (1970).

\bibitem{Payne}  R. Payne and I.E. Theodorou, J. Phys. Chem. {\bf 76}, 2892
(1972).

\bibitem{Masood}  A.K.M. Masood, R.A. Pethrik, A.J. Barlow, M.G. Kim, R.P.
Plowiec, D. Barraclough, and J.A. Ladd, Adv. Mol. Relax. Process. {\bf 9},
29 (1976).

\bibitem{Huck}  J.R. Huck, G.A. Noyel, L.J. Jorat, and A.M. Bondeau, Journal
of Electrostatics {\bf 12}, 221 (1982).

\bibitem{Barthel}  J. Barthel, K. Bachhuber, E. Buchner, J.B. Gill, and M.
Kleebauer, Chem. Phys. Lett. {\bf 167}, 62 (1990).

\bibitem{Schoen}  A. Sch\"{o}nhals, F. Kremer, A. Hofmann, E.W. Fischer, and
E. Schlosser, Phys. Rev. Lett. {\bf 70}, 3459 (1993); A. Sch\"{o}nhals, F.
Kremer, A. Hofmann, and E.W. Fischer, Physica A {\bf 201}, 263 (1993).

\bibitem{BohmPC}  R. B\"{o}hmer, B. Schiener, J. Hemberger, and R.V.
Chamberlin, Z. Phys. B {\bf 99}, 91 (1995).

\bibitem{Le}  R.L. Leheny and S.R. Nagel, Europhys. Lett. {\bf 39}, 447
(1997).

\bibitem{CD}  D.W. Davidson and R.H. Cole, J. Chem. Phys. {\bf 19}, 1484
(1951).

\bibitem{KWW}  R. Kohlrausch, Ann. Phys. {\bf 167}, 179 (1854); G. Williams
and D.C. Watts, Trans. Faraday Soc. {\bf 66}, 80 (1970).

\bibitem{Brand}  R. Brand, P. Lunkenheimer, and A. Loidl, Phys. Rev. B {\bf %
56}, R5713 (1997).

\bibitem{Boett}  C.J.F. B\"{o}ttcher and P. Bordewijk, {\it Theory of
Electric Polarization, Vol. II} (Elsevier, Amsterdam, 1978).

\bibitem{Stick2}  F. Stickel, E.W. Fischer, and R. Richert, J. Chem. Phys. 
{\bf 104}, 2043 (1996).

\bibitem{VFT}  H. Vogel, Z. Phys. {\bf 22}, 645 (1921); G.S. Fulcher, J. Am.
Ceram. Soc. {\bf 8}, 339 (1925); G. Tammann and W. Hesse, Z. Anorg. Allg.
Chem. {\bf 156}, 245 (1926).

\bibitem{Robdip}  P. Lunkenheimer, R. Brand, U. Schneider, and A. Loidl,
submitted to {\it Proceedings of the 8th Tohwa University International
Symposium on Slow Dynamics in Complex Systems} (AIP Conference Proceedings,
New York, 1999).

\bibitem{Wahn}  G. Wahnstr\"{o}hm and L.J. Lewis, Progr. Theor. Phys. Suppl. 
{\bf 126}, 261 (1997).

\bibitem{Schsim}  S. K\"{a}mmerer, W. Kob, and R. Schilling, Phys. Rev. E 
{\bf 58}, 2141 (1998).

\bibitem{Ohl}  J. Wuttke, M. Ohl {\it et al.}, to be published.

\bibitem{AG}  G. Adam and J.H. Gibbs, J. Chem. Phys. {\bf 43}, 139 (1965).

\bibitem{FV}  M.H. Cohen and D. Turnbull, J. Chem. Phys. {\bf 31}, 1164
(1959).

\bibitem{cumcom}  H.Z. Cummins, J. Hernandez, W.M. Du, and G. Li, Phys. Rev.
Lett. {\bf 73}, 2935 (1994).

\bibitem{Stick1}  F. Stickel, E.W. Fischer, and R. Richert, J. Chem. Phys. 
{\bf 102}, 6251 (1995).

\bibitem{EFV}  M.H. Cohen and G.S. Grest, Phys. Rev. B {\bf 20}, 1077
(1979); G.S. Grest and M.H. Cohen, Adv. Chem. Phys. {\bf 48}, 455 (1981).

\bibitem{Schpriv}  R. Schilling, private communication.

\bibitem{Kiv96}  D. Kivelson, G. Tarjus, X. Zhao, and S.A. Kivelson, Phys.
Rev. E {\bf 53}, 751 (1996).

\bibitem{remschoen}  These authors compared $T\Delta \varepsilon (T)$ with
MCT, presumably in order to correct for dipolar interactions dominating
their $\Delta \varepsilon (T)$, in contrast to the findings in the present
work.

\bibitem{remwing}  In a recent investigation \cite{wing} we have shown that
the excess wing is not universally present in {\it orientationally}
disordered materials, i.e. it seems to be a property inherent to {\it %
structural} glass-formers.

\bibitem{Kudphen}  A. Kudlik, Ph.D. Thesis, Universit\"{a}t Bayreuth, 1997.

\bibitem{Chambrem}  However, a recent extension of the model may enable a
description of the complete spectra even including the minimum region (R.V.
Chamberlin, preprint).

\bibitem{Schoen2}  A. Sch\"{o}nhals, F. Kremer, and E. Schlosser, Phys. Rev.
Lett. {\bf 67}, 999 (1991).

\bibitem{Dendz}  Z. Dendzik, M. Paluch, Z. Gburski, and J. Ziolo, J. Phys.:
Condens. Matter {\bf 9}, L339 (1997).

\bibitem{Kud}  A. Kudlik, S. Benkhof, R. Lenk, and E. R\"{o}ssler, Europhys.
Lett. {\bf 32}, 511 (1995).

\bibitem{Men}  N. Menon, K.P. O%
%TCIMACRO{\UNICODE{0xb4}}%
%BeginExpansion
\'{}%
%EndExpansion
Brien, P.K. Dixon, L. Wu, S.R. Nagel, B.D. Williams, and J.P. Carini, J.
Non-Cryst. Solids {\bf 141}, 61 (1992).

\bibitem{LP}  D.L. Leslie-Pelecky and N.O. Birge, Phys. Rev. Lett. {\bf 72},
1232. (1994).

\bibitem{cousin}  K.L. Ngai, Phys. Rev. E {\bf 57}, 7346 (1998); K.L. Ngai,
private communication.

\bibitem{Strom}  U. Strom, J.R. Hendrickson, R.J. Wagner, and P.C. Taylor,
Solid State Commun. {\bf 15}, 1871 (1974); U. Strom and P.C. Taylor, Phys.
Rev. B {\bf 16}, 5512 (1977); C. Liu and C.A. Angell, J. Chem. Phys. {\bf 93}%
, 7378 (1990).

\bibitem{remadd}  One has to be aware that the assumption of an additive
superposition of different contributions to the susceptibility may be
oversimplified, [for an alternative see, e.g., J. Colmenero, A. Arbe, and A.
Alegria, Phys. Rev. Lett. {\bf 71}, 2603 (1993)] but at least it seems
sufficient for a preliminary analysis.

\bibitem{Ngaipriv}  K.L. Ngai, private communication.

\bibitem{colo}  W.K. Lee, H.F. Liu, and A.S. Nowick, Phys. Rev. Lett. {\bf 67%
}, 1559 (1991); D.L. Sidebottom, P.F. Green, and R.K. Brow, Phys. Rev. Lett. 
{\bf 74}, 5068 (1995); C. Cramer, K. Funke, and T. Saatkamp, Philos. Mag. B 
{\bf 71}, 701 (1995).

\bibitem{coloexpl}  R.H. Cole and E. Tombari, J. Non-Cryst. Solids {\bf %
131-133}, 969 (1991); K.L. Ngai, U. Strom, and O. Kanert, Phys. Chem.
Glasses {\bf 33},\ 109 (1992); B.S. Lim, A.V. Vaysleyb, and A.S. Nowick,
Appl. Phys. A {\bf 56}, 8 (1993); S.R. Elliott, Solid State Ionics {\bf 70/71%
}, 27 (1994);\ D.L. Sidebottom, P.F. Green, and R.K. Brow, J. Non-Cryst.
Solids {\bf 203}, 300 (1996); K.L. Ngai, H. Jain, and O. Kanert, J.
Non-Cryst. Solids {\bf 222}, 383 (1997).

\bibitem{fukglas}  P. Lunkenheimer, U. Schneider, R. Brand, and A. Loidl,
submitted to {\it Proceedings of the 8th Tohwa University International
Symposium on Slow Dynamics in Complex Systems} (AIP Conference Proceedings,
New York, 1999).

\bibitem{Novi}  V.N. Novikov, Phys. Rev. B {\bf 58}, 8367 (1998).

\bibitem{Berg}  J. Ma, D. Vanden Bout, and M. Berg, Phys. Rev. E {\bf 54},
2786 (1996).

\bibitem{Borj}  L. B\"{o}rjesson, M. Elmroth, and L.M. Torell, Chem. Phys. 
{\bf 149}, 209 (1990).

\bibitem{Gotzepriv}  W. G\"{o}tze, private communication.

\bibitem{Leb}  M.J. Lebon, C. Dreyfus, Y. Guissani, R.M. Pick, and H.Z.
Cummins, Z. Phys. B {\bf 103}, 433 (1997).

\bibitem{Schkug}  R. Schilling and T. Scheidsteger, Phys. Rev. E {\bf 56},
2932 (1997).

\bibitem{soft}  V.G. Karpov, M.I. Klinger, and F.N. Ignatiev, Sov. Phys.
JETP {\bf 57}, 439 (1983); U. Buchenau, Yu.M. Galperin, V.L. Gurevich, D.A.
Parshin, M.A. Ramos, and H.R. Schober, Phys. Rev. B {\bf 46}, 2798 (1992).

\bibitem{phonloc}  V.K. Malinovsky, V.N. Novikov, and A.P. Sokolov, J.
Non-Cryst. Solids {\bf 485} (1987); S.R. Elliott, Europhys. Lett {\bf 19},
201 (1992).

\bibitem{Schirm}  W. Schirmacher, G. Diezemann, and C. Ganter, Phys. Rev.
Lett. {\bf 81}, 136 (1998).

\bibitem{Fra}  T. Franosch, W. G\"{o}tze, M.R. Mayr, and A.P. Singh, Phys.
Rev. E {\bf 55}, 3183 (1997).
\end{references}
\end{document}